\documentclass[11pt, prd, aps, showpacs, nofootinbib, superscriptaddress]{revtex4-1}

\usepackage{amsmath,amssymb}
\usepackage{graphicx}
\usepackage[many]{tcolorbox}

\usepackage[normalem]{ulem}

\newcommand{\be}{\begin{equation}}
\newcommand{\beqn}{\begin{eqnarray}}
\newcommand{\ee}{\end{equation}}
\newcommand{\eeqn}{\end{eqnarray}}
\newcommand{\bea}{\begin{align}}
\newcommand{\eea}{\end{align}}
\newcommand{\nn}{\nonumber}
\allowdisplaybreaks

\newcommand{\Maconn}{\raisebox{-0.43\totalheight}{\includegraphics[scale=0.60]{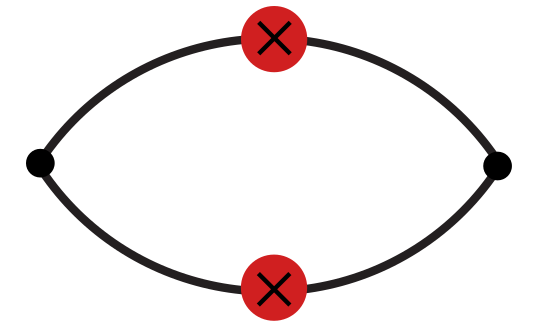}}}
\newcommand{\Madisc}{\raisebox{-0.43\totalheight}{\includegraphics[scale=0.60]{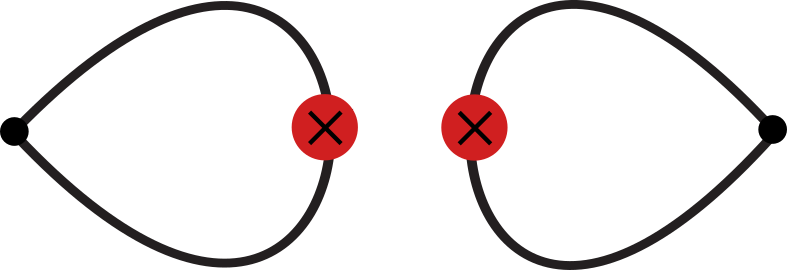}}}
\newcommand{\Maconnpp}{\raisebox{-0.42\totalheight}{\includegraphics[scale=0.60]{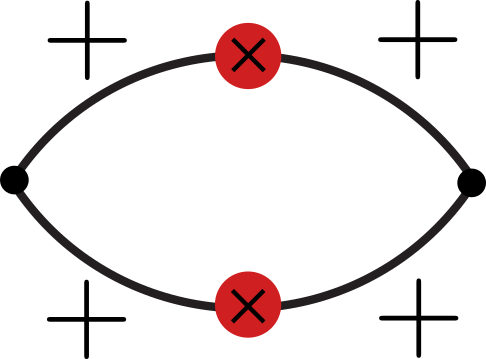}}}
\newcommand{\Madiscpp}{\raisebox{-0.45\totalheight}{\includegraphics[scale=0.60]{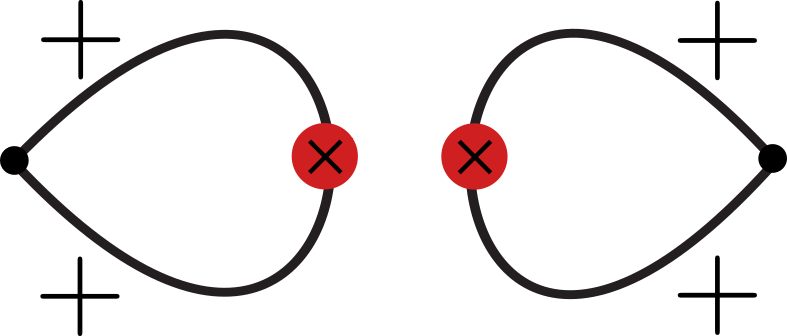}}}
\newcommand{\Maconnpm}{\raisebox{-0.42\totalheight}{\includegraphics[scale=0.60]{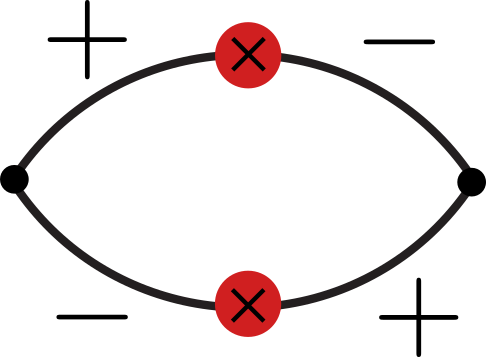}}}
\newcommand{\Madiscpm}{\raisebox{-0.45\totalheight}{\includegraphics[scale=0.60]{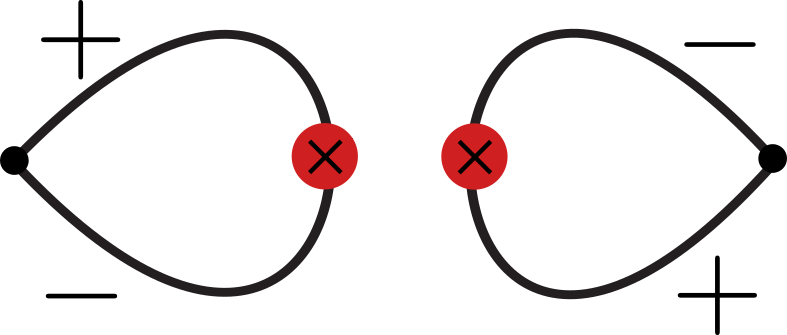}}}
\newcommand{\Mcconn}{\raisebox{-0.43\totalheight}{\includegraphics[scale=0.50]{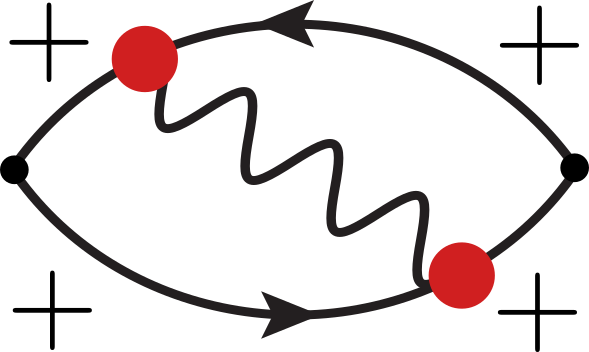}}}
\newcommand{\Mcdisc}{\raisebox{-0.23\totalheight}{\includegraphics[scale=0.50]{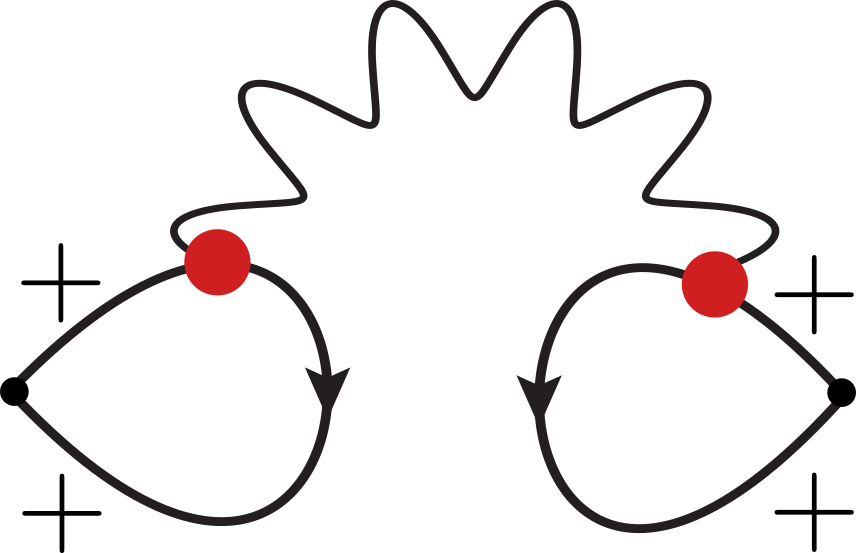}}}
\newcommand{\Mcconnpm}{\raisebox{-0.43\totalheight}{\includegraphics[scale=0.50]{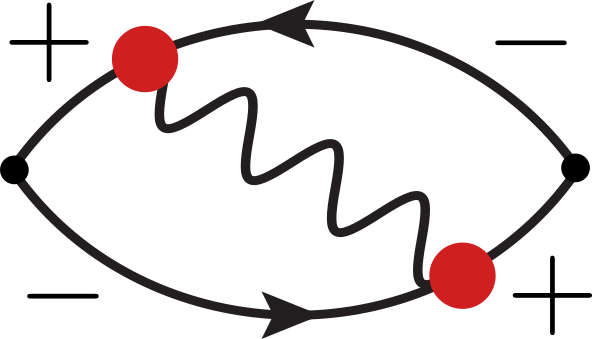}}}
\newcommand{\Mcdiscpm}{\raisebox{-0.23\totalheight}{\includegraphics[scale=0.50]{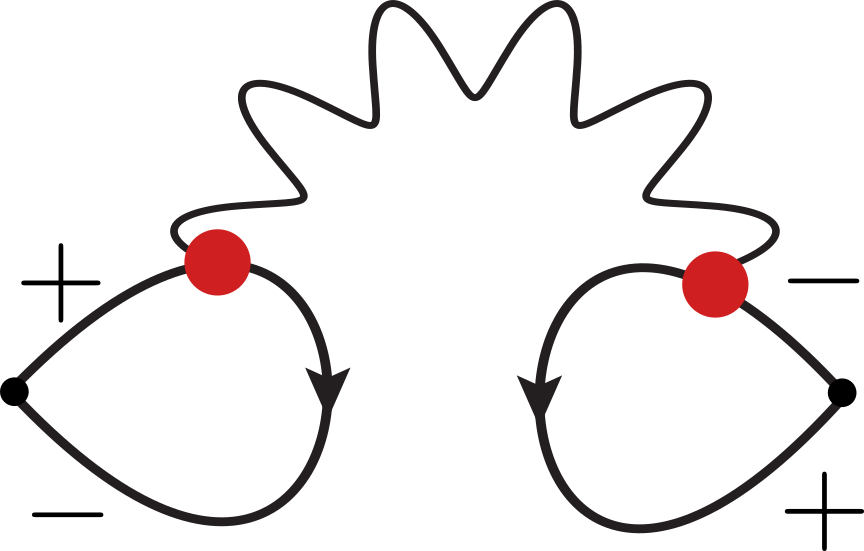}}}

\newcommand{\Isocorr}{\raisebox{-0.43\totalheight}{\includegraphics[scale=0.45]{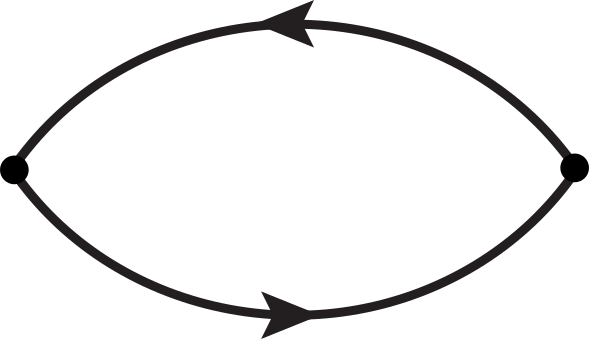}}}
\newcommand{\Mcconnnor}{\raisebox{-0.43\totalheight}{\includegraphics[scale=0.45]{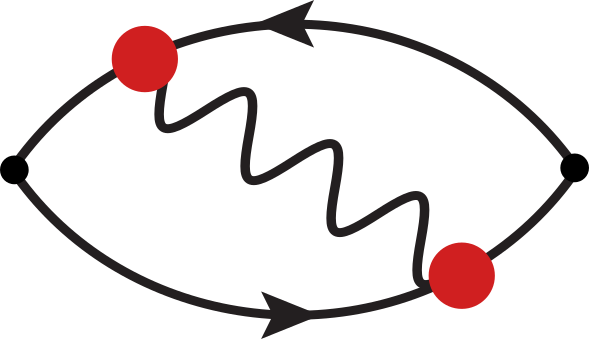}}}
\newcommand{\Mcdiscnor}{\raisebox{-0.23\totalheight}{\includegraphics[scale=0.45]{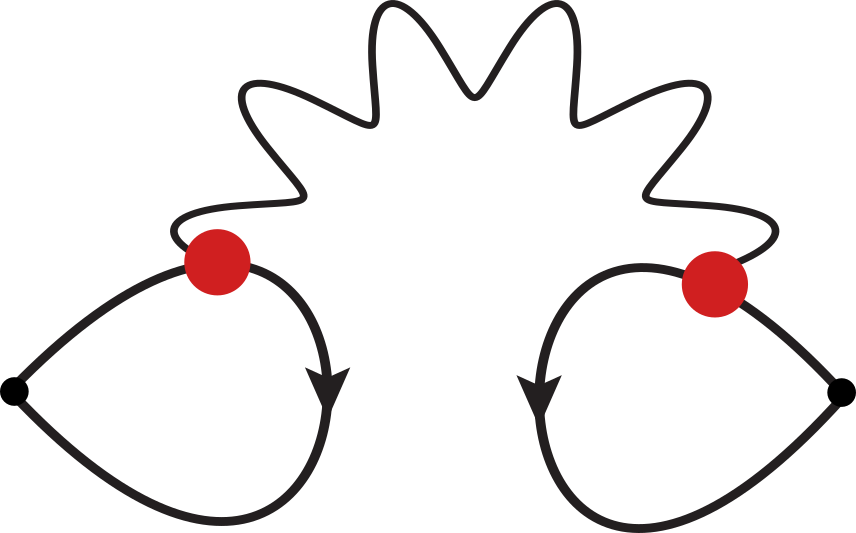}}}

\newcommand{\Romatre}{Dipartimento di Fisica, Universit\`a  Roma Tre and INFN, Sezione di Roma Tre,\\ Via della Vasca Navale 84, I-00146 Rome, Italy}
\newcommand{\RomatreINFN}{Istituto Nazionale di Fisica Nucleare, Sezione di Roma Tre,\\ Via della Vasca Navale 84, I-00146 Rome, Italy}
\newcommand{\Romadue}{Dipartimento di Fisica and INFN, Universit\`a di Roma ``Tor Vergata",\\ Via della Ricerca Scientifica 1, I-00133 Roma, Italy}

\begin{document}

\title{\Large Rotated twisted-mass: a convenient regularization scheme \\ for isospin breaking QCD and QED lattice calculations}

\author{R.\,Frezzotti}\affiliation{\Romadue} 
\author{G.\,Gagliardi}\affiliation{\RomatreINFN}
\author{V.\,Lubicz}\affiliation{\Romatre} 
\author{F.\,Sanfilippo}\affiliation{\RomatreINFN}
\author{S.\,Simula}\affiliation{\RomatreINFN}

\begin{abstract}
\vspace{0.5cm}
We propose a scheme of lattice twisted-mass fermion regularization which is particularly convenient for application to isospin breaking (IB) QCD and QED calculations, based in particular on the so called RM123 approach, in which the IB terms of the action are treated as a perturbation. The main, practical advantage of this scheme is that it allows the calculation of IB effects on some mesonic observables, like e.g. the $\pi^+ - \pi^0$ mass splitting, using lattice correlation functions in which the quark and antiquark fields in the meson are regularized with opposite values of the Wilson parameter $r$. These correlation functions are found to be affected by much smaller statistical fluctuations, with respect to the analogous functions in which quark and antiquark fields are regularized with the same value of $r$. Two numerical application of this scheme, that we call \textit{rotated twisted-mass}, within pure QCD and QCD+QED respectively, are also provided for illustration. 
\end{abstract}

\maketitle

\section{Introduction}

The evaluation of isospin breaking (IB) effects in hadronic observables has become, in recent years, an important goal of lattice QCD and QED calculations in flavor physics~\cite{FLAG}. The reason is that, while the breaking of isospin symmetry in Nature, which is due to both the mass and electric charge difference of the up and down quarks, is expected to be small, i.e. at the level of 1\%, the remarkable improving of the experimental and theoretical precision is such that, for several phenomenological quantities of interest, IB effects are no longer negligible. As an example, we mention here the determination of the CKM matrix elements $V_{ud}$ and $V_{us}$ from leptonic and semileptonic kaon and pion decays. For these processes, an accuracy at the level of few per mille has been reached by both the experimental measurements of the relevant decay rates and by the lattice determinations of the corresponding hadronic parameters in the isospin symmetric limit, namely the ratio of leptonic decay constant $f_K/f_\pi$ and the semileptonic form factor $f_+^{K\pi}(0)$~\cite{FLAG}. It is then clear that, for these processes, IB effects have to be taken into account in lattice calculations.

Among the various lattice regularization which are commonly employed in lattice simulations, we are concerned in this paper with the twisted-mass (TM) regularization of the fermionic action~\cite{TM,FR1}. The main advantage of TM fermions is that ${\cal O}(a)$-improvement is automatically guaranteed for parity-conserving physical observables~\cite{FR1}, while the numerical cost of the simulation remains relatively small. In the original TM action for lattice QCD~\cite{TM}, the Wilson term has a Dirac structure proportional to $\gamma_5$, in the so called physical basis. Moreover, the up and down quarks are regularized with opposite values of the Wilson parameter, i.e. $r_u = +1$ and $r_d = -1$, that is the Wilson term is proportional to matrix $\tau_3$ in the isospin space.

The issue of how to include IB effects in lattice QCD calculations has been addressed by the lattice community using a variety of different methods, and remarkably accurate results, using different lattice discretizations, have been obtained (see e.g. Refs.~\cite{Blum:2010ym,Ishikawa:2012ix,Aoki:2012st, Borsanyi:2014jba, Endres:2015gda, Horsley:2015eaa, Fodor:2016bgu, Boyle:2016lbc, Borsanyi:2020mff}). In particular, the TM regularization has been largely applied in the last years to the calculation of several observables, including hadronic masses~\cite{IBs,IBem,IBmasses}, leptonic decay rates~\cite{IBrates,IBrates2,IBrates3,IBrates4} and the hadronic vacuum polarization contribution to the muon anomalous magnetic moment $a_\mu$~\cite{IBg-2,IBg-21,IBg-22}. In all these calculations, IB effects have been evaluated using the so called RM123 approach introduced in Refs.~\cite{IBs,IBem}. The basic idea of this method is that the IB term in the QCD action, which is proportional to up-down quark mass difference $\Delta m=m_d-m_u$, as well as the QED interaction term of quarks, which is proportional to the electromagnetic coupling $\alpha_{em}$, are treated as small perturbations and expanded to the desired order. In most cases of interest, keeping only the leading terms in the expansion, i.e. terms of ${\cal O}(\Delta m)$ and ${\cal O}(\alpha_{em})$, is by far sufficient, since the corrections are expected to be of $\Delta m/\Lambda_{QCD} \sim 1\%$ and $\alpha_{em} \sim 1\%$. For this reason, all phenomenological applications so far have been limited to this case. In this paper, however, we will present as a numerical application, for the first time, a calculation of the strong IB effect at ${\cal O}(\Delta m^2)$.

The aim of the present paper is to discuss a scheme for TM regularization of lattice QCD, that we call \textit{rotated twisted-mass} (RTM), which is particularly convenient for lattice calculations of IB effects, in particular within the RM123 approach discussed above. The main, practical advantage of this scheme is that it allows the calculation of IB effects on some mesonic observables, like e.g. the $\pi^+ - \pi^0$ mass splitting, using lattice correlation functions in which the quark and antiquark fields in the meson are regularized, in the TM setup, with opposite values of the Wilson parameter $r$, i.e. $r=\pm 1$. This is at variance with the standard TM regularization for light quarks, in which, for instance, the 2-point function of the neutral pion, being the meson composed by a quark and an antiquark with the same flavor, either $u$ or $d$, is regularized with the same value of the Wilson parameter $r$. We will show in this paper that mesonic correlation functions in which quark and antiquark fields are regularized with opposite values of the Wilson parameter are affected by substantially smaller statistical fluctuations, with respect to the corresponding functions in which quark and antiquark fields are regularized with the same value of $r$. This advantage turns out to be especially relevant in the calculation of disconnected quark diagrams, which are known to be particularly noisy from the statistical point of view and, therefore, computationally very expensive.  

In order to illustrate the advantage of RTM scheme, we will present in this paper two numerical applications, namely the calculations of the $\pi^+ - \pi^0$ mass splitting in pure QCD, at ${\cal O}(\Delta m^2)$, and in QCD+QED, at ${\cal O}(\alpha_{em})$. These two example are particularly suitable for the illustration of the RTM scheme, because they both involve the calculation of a connected and a disconnected quark diagram. On the other hand, since these examples are only provided here for illustrative purposes, we will present numerical results obtained only at a fixed value of quark masses and lattice spacing, and postpone the more complete and phenomenologically interesting calculations to future studies.

The plan of the remaining of this paper is the following. In Sect.\,1, we will consider the case of pure QCD and, after summarizing the basic ingredients of the RM123 approach for the calculations of IB effects, extended up to the second order in $\Delta m$, we will illustrate our proposal of the RTM scheme, and present a numerical application to the calculation of the $\pi^+ - \pi^0$ mass splitting. In Sect.\,2, we will extend the RTM scheme to QED, and present its application to the calculation of the $\pi^+ - \pi^0$ mass splitting at ${\cal O}(\alpha_{em})$. We end this paper by presenting some final considerations in the Conclusions.

\section{RTM scheme and QCD isospin breaking corrections}

In this section, we illustrate the RTM scheme by considering its implementation in pure QCD, applied to the  study of IB effects induced by the quark mass difference $\Delta m=m_d-m_u$. The extension of the  RTM scheme to QED and the calculation of electromagnetic IB corrections will be addressed in the next section. 

We start the discussion by reviewing the basic ingredients of the RM123 approach for evaluating IB corrections in pure QCD~\cite{IBs}, that we extend here up to the second order in $\Delta m$. 

By having in mind the application of the method with different choices of the lattice regularization, specifically either standard or rotated TM, we do not specify the regularization from the very beginning, and adopt for the QCD fermionic action a simple continuum notation. We are assuming, implicitly, that some kind of lattice regularization has been implemented for the quark fields. In addition, throughout this paper, we will only limit the discussion to the theory with the $u$ and $d$ quarks only, since the extension of the RTM regularization to other quark doublets is just straightforward.

The RM123 approach relies on treating the IB term in the QCD action, which is proportional to the mass difference $\Delta m=m_d-m_u$, as a small perturbation. The QCD Lagrangian for the up and down quarks has the form
\be
\label{LQCD}
{\cal L}_{QCD} = {\cal L}_{kin} + {\cal L}_m \, ,
\ee
where ${\cal L}_{kin}$ is the kinetic term for massless quark,
\be
\label{Lkin}
{\cal L}_{kin} = \bar u\, \gamma_\mu D_\mu u + \bar d\, \gamma_\mu D_\mu d = \bar Q\, \gamma_\mu D_\mu Q \, ,
\ee
with $Q=(u,d)$, and ${\cal L}_m$ is the quark mass term. The latter can be written as the sum of a term which is $SU(2)$ symmetric plus a term which violates the isospin symmetry:
\begin{align}
\label{Lm}
	{\cal L}_m & = m_u \, \bar u u + m_d \, \bar d d =  
	\frac{m_u + m_d}{2}\, (\bar u u + \bar d d ) - \frac{m_d - m_u}{2}  \, (\bar u u - \bar d d ) = \nn \\
	& = m\, (\bar u u + \bar d d ) - \Delta m  \, (\bar u u - \bar d d ) = 
	m\, \bar Q Q - \Delta m  \, \bar Q \tau_3 Q \, ,
\end{align}
where $m$ and $\Delta m$ are given by
\be
m = \frac{1}{2}\, (m_u + m_d) \qquad , \qquad \Delta m = \frac{1}{2}\, (m_d - m_u) \, .
\ee

For later discussion, we find convenient to rewrite the Lagrangian \eqref{LQCD} as the sum of the
isospin symmetric part ${\cal L}_0$ and the IB contribution ${\cal L}_{IB}$, i.e.
\be
{\cal L}_{QCD} = {\cal L}_0 + {\cal L}_{IB} \, ,
\ee
where
\be
\label{L0}
{\cal L}_0 =  \bar Q \left( \gamma_\mu D_\mu + m \right) Q
\ee
and 
\be
\label{LIB}
{\cal L}_{IB} =   - \Delta m  \, \bar Q \tau_3 Q \, .
\ee
Consequently, the only IB term in the action is $S_{IB} = - \Delta m  \, \hat{\cal S}$ where $\hat{\cal S}$ is the isospin diagonal scalar operator
\be
\label{hatS}
\hat{\cal S} = \sum_x (\bar Q \tau_3 Q)(x) = \sum_x (\bar u u - \bar d d )(x)\, .
\ee

In the RM123 approach, $S_{IB}$ is  treated as a perturbation and expanded in the exponential of the path-integral. If we carry out the expansion up to the second order, we find that the vacuum expectation value of a generic operator ${\cal O}$ is given by
\begin{align}
\label{2ndorder}
	\langle {\cal O} \rangle & = \frac{\int D\phi\, {\cal O}\, e^{-(S_0 - \Delta m \, \hat{\cal S})}}{\int D\phi \, e^{-(S_0 - \Delta m \, \hat{\cal S})}} \simeq
	\frac{\int D\phi\, {\cal O}\, ( 1 + \Delta m \, \hat{\cal S} + \frac{1}{2} \Delta m^2 \, \hat{\cal S}^2 ) \, e^{-S_0}}{\int D\phi \, ( 1 + \Delta m \, \hat{\cal S} + \frac{1}{2} \Delta m^2 \, \hat{\cal S}^2 ) \, e^{-S_0}} = \nn \\[0.4em]
	& = \frac{\langle {\cal O} \rangle_0 +  \Delta m \, \langle {\cal O}\, \hat{\cal S} \rangle_0 + \frac{1}{2} \Delta m^2 \, \langle {\cal O}\, \hat{\cal S}^2 \rangle_0}{1 +  \Delta m \, \langle \hat{\cal S} \rangle_0 + \frac{1}{2} \Delta m^2 \, \langle \hat{\cal S}^2 \rangle_0} \simeq \nn \\[0.4em]
	& \simeq \langle {\cal O} \rangle_0 +  \Delta m \, \langle {\cal O}\, \hat{\cal S} \rangle_0 + \frac{1}{2} \Delta m^2 \, \left( \langle {\cal O}\, \hat{\cal S}^2 \rangle_0 - \langle {\cal O} \rangle_0 \langle \hat{\cal S}^2 \rangle_0 \right) \, ,
\end{align}
where $S_0$ is the isospin symmetric part of the QCD action, which now also include the pure gauge part of the action, and $\langle \cdot \rangle_0$ represents the vacuum expectation value in the iso-symmetric theory. In Eq.\,\eqref{2ndorder}, we have exploited that $\langle \hat{\cal S} \rangle_0=0$  due to isospin symmetry.

Let us consider, as an example, the application of the method to the calculation of the $\pi^+ - \pi^0$ pion mass splitting. Using Eq.\,\eqref{2ndorder}, one finds that the relevant correlation function for computing this splitting, i.e. the difference between the neutral and charged pion propagators, vanishes both at zero and first order in $\Delta m$, and it is given at second order by \cite{IBs}
\be
\label{pippi0}
C_{\pi^0\pi^0}-C_{\pi^+\pi^+} = 2\, \Delta m^2 \left[ \,\,\Maconn \,\,\,\,-\,\,\,\,\Madisc \,\, \right] \, ,
\ee
where the crosses on the quark lines in the Feynman diagrams denote the insertion of the scalar operator $\hat{\cal S}$ of Eq.\,\eqref{hatS}.

We now specify the lattice regularization of the fermionic action. 

In the standard TM regularization, at maximal twist, the Lagrangian for the light quark doublet $Q=(u,d)$, in the so-called physical basis, has the form~\cite{TM,FR1}
\be
\label{TMaction}
{\cal L}_{TM} =  \bar Q \left[ \gamma_\mu \widetilde \nabla_\mu - i \gamma_5 \tau_3 \, W(m_{cr}) + m \right] Q -  \Delta m\, \bar Q\, \tau_3Q \, ,
\ee
where $\widetilde \nabla_\mu$ is the lattice symmetric covariant derivative, written in terms of the forward $(\nabla_\mu )$ and backward ($ \nabla^*_\mu$) covariant derivatives, 
\be
\widetilde \nabla_\mu = \frac{1}{2} \left( \nabla^*_\mu + \nabla_\mu \right)
\ee
and $W(m_{cr})$ is the critical Wilson term, which includes the mass and is globally odd under $r \to -r$,
\be
W(m_{cr}) = - a\, \frac{r}{2}\, \nabla_\mu \nabla^*_\mu + m_{cr}(r)\, .
\ee

A characteristic feature of the TM action \eqref{TMaction} is that, due to the presence of the isospin $\tau_3$ matrix in the Wilson term, the up and down quark fields are regularized with opposite values of the Wilson parameter, i.e. $r_u = +1$ and $r_d = -1$. Since the IB term in  \eqref{TMaction} is also proportional to $\tau_3$, and it is therefore diagonal in flavor space, one finds that in the disconnected quark diagrams of Eq.\,\eqref{pippi0}, which enter the correlation function of the neutral pion $C_{\pi^0\pi^0}$, the initial and final states must be composed by quarks regularized with the same value of the Wilson parameter. That is, by specifying the sign of the Wilson parameter in the diagrams, the correlation function to be actually computed with the TM action is\footnote{Strictly speaking, in the TM regularization one finds that the first diagram in Eq.\,\eqref{pippi0TM} is actually the average of two contributions, coming from the neutral and charged pion respectively, one with equal and one with opposite values of the Wilson parameters in the two quark lines. The two contributions differ by terms of ${\cal O}(a^2)$, so that any choice for them is equally legitimate. In the second diagram of  Eq.\,\eqref{pippi0TM}, however, the values of the Wilson parameter in each ``loop'' are necessarily equal.}
\be
\label{pippi0TM}
C_{\pi^0\pi^0}-C_{\pi^+\pi^+} = 2\, \Delta m^2 \left[ \,\, \Maconnpp \,\,\,\,- \,\,\,\,\Madiscpp \,\, \right] \, ,
\ee
This is unfortunate, because mesonic correlators composed by quarks regularized with equal values of $r$ are statistically much noisier than the corresponding correlators composed by quarks regularized with opposite values of $r$. Therefore, with the aim of improving the statistical accuracy of the calculation, we propose to adopt the RTM scheme that we now proceed to illustrate.

The RTM scheme relies on the introduction of the following rotated basis for the quark fields
\be
\label{qrotated}
Q' = \left(\begin{array}{cc} u' \\ d' \end{array} \right) = U Q =
\frac{1}{\sqrt{2}}  \left(\begin{array}{cc} 1 & 1 \\ -1 & 1 \end{array} \right) \left(\begin{array}{cc} u \\ d \end{array} \right) =
\frac{1}{\sqrt{2}} \left(\begin{array}{cc} u +d \\ d-u \end{array} \right)\, .
\ee

In terms of these fields, the isospin symmetric part of the Lagrangian, given in Eq.\,\eqref{L0}, being a scalar in flavor space, is invariant, i.e.
\be
\label{L0rotated}
{\cal L}^{\prime}_{0} = \bar Q' \left( \gamma_\mu D_\mu + m \right) Q' \, ,
\ee
while the IB term \eqref{LIB}, proportional to $\Delta m$, is now rotated in the direction of $\tau_1$ in the isospin space,
\be
\label{Lmrotated}
	{\cal L}^{\prime}_{IB}  = + \Delta m  \, \bar Q' \tau_1 Q = + \Delta m  \, (\bar u' d' + \bar d' u' ) \, .
\ee

Clearly, the Lagrangian in the new basis describes the same theory, since the physical content of the theory is not changed by a rotation of the quark fields. In particular, the mass eigenstates of the theory are always composed by the $(u,d)$ fields, since the mass term of the Lagrangian is diagonal in that basis.

We now introduce, however, the TM lattice regularization, and opt for writing the Wilson term diagonal in the rotated basis, i.e. we regularize the $(u',d')$ quarks with $r_{u'} = +1$ and $r_{d'} = -1$. In this way, we obtain the lattice Lagrangian in the form
\be
\label{RTMaction}
{\cal L}_{RTM} =  \bar Q' \left[ \gamma_\mu \widetilde \nabla_\mu - i \gamma_5 \tau_3 \, W(m_{cr}) + m \right] Q' +  \Delta m\, \bar Q' \, \tau_1 Q' \, .
\ee
This Lagrangian is no longer equivalent to the TM Lagrangian \eqref{TMaction}, since rotating back to the physical basis of the $Q=(u,d)$ fields one finds
\be
\label{RTMactionphys}
{\cal L}_{RTM}^{(Q)} =  \bar Q \left[ \gamma_\mu \widetilde \nabla_\mu - i \gamma_5 \tau_1 \, W(m_{cr}) + m \right] Q -  \Delta m\, \bar Q \, \tau_3 Q \, ,
\ee
which differs from Eq.\,\eqref{TMaction} for the direction of the Wilson term in flavor space. We refer to the regularization scheme of Eq.\,\eqref{RTMaction} or \eqref{RTMactionphys} as the RTM scheme.

We note that the RTM Lagrangian \eqref{RTMactionphys}, in the physical basis, coincides with the TM regularization proposed in Ref.\,\cite{NdegTM} for TM simulations of non-degenerate quarks, and adopted by the ETM Collaborations for the strange and charm doublet. At the time it was introduced, this was motivated by the fact that the Dirac operator for the action \eqref{TMaction} is complex for $\Delta m \neq 0$ and thus the action cannot be used in numerical simulations of the $N_f=2+1+1$ theory. Here, however, we have a different perspective. In the framework of the RM123 method for evaluating IB corrections, we envisage to exclude the IB term from the action and to treat it as a perturbation. Therefore, the reality and positivity of the fermionic determinant is not an issue.
The action with which the vacuum expectation values $\langle \cdot \rangle_0$ of Eq.\,\eqref{2ndorder} are calculated is exactly the same in the RTM scheme of Eq.\,\eqref{RTMaction} and in the standard TM scheme of Eq.\,\eqref{TMaction}. It also follows, from this consideration, that all the features characterizing the standard TM regularization remains valid in the RTM scheme, including primarily the non-perturbative improvement at ${\cal O}(a^2)$ of parity conserving observables.

The only difference between the RTM and standard TM regularizations, within the RM123 approach, comes from the insertions of the IB operator proportional to $\Delta m$, which is proportional to $\tau_1$ rather than $\tau_3$ in flavor space. Therefore, in particular, in the RTM scheme the insertion of the mass term proportional to $\Delta m$ also induces a flipping of the Wilson parameter $r$. We are now going to show that this feature leads, for some interesting IB mesonic observables, to a significantly improved statistical accuracy.

In order to identify the physical content of the correlation functions in the RTM scheme, we have to express the meson fields in the rotated basis in terms of the physical mesons, using Eq.\,\eqref{qrotated} to relate the quark fields in the two basis. In particular, one finds that the ``neutral" and ``charged" rotated pion fields are given by
\begin{align}
\label{pirotated}
& P_{\pi'^0} = \frac{1}{\sqrt{2}}\, (\bar u' \gamma_5 u' - \bar d' \gamma_5 d' ) = \frac{1}{\sqrt{2}}\, (\bar d \gamma_5 u + \bar u \gamma_5 d ) =
\frac{1}{\sqrt{2}} \left(P_{\pi^+} + P_{\pi^-} \right) \\
& P_{\pi'^+} = \bar d' \gamma_5 u' = \frac{1}{2}\, (-\bar u \gamma_5 u + \bar d \gamma_5 d + \bar d \gamma_5 u - \bar u \gamma_5 d ) =\,   \frac{1}{2} \left( P_{\pi^+} - P_{\pi^-} \right) - \frac{1}{\sqrt{2}}\,P_{\pi^0} \nn \\
& P_{\pi'^-} = \bar u' \gamma_5 d' = \frac{1}{2}\, (-\bar u \gamma_5 u + \bar d \gamma_5 d - \bar d \gamma_5 u + \bar u \gamma_5 d ) =  \,  - \frac{1}{2} \left( P_{\pi^+} - P_{\pi^-} \right) - \frac{1}{\sqrt{2}}\,P_{\pi^0}  \nn
\end{align}

The advantage of the RTM regularization for the calculation of selected observables is then easily illustrated in the case of the $\pi^+ - \pi^0$ pion splitting. Let us consider the correlation function $C_{\pi'^+\pi'^-}$, which describes, in the rotated basis, the mixing between the positive and negative rotated pion fields. This mixing is absent in the iso-symmetric theory, and it is generated at second order by a double insertion of the isospin breaking correction proportional to $\Delta m$, since this correction is flavor-changing in the RTM scheme. Using Eqs.\,\eqref{pirotated}, we find that indeed $C_{\pi'^+\pi'^-}$ is simply related to the difference of the neutral and charged (physical) pion propagators:
\be
\label{cpippim}
C_{\pi'^+\pi'^-} = \frac{1}{2}\, C_{\pi^0\pi^0}  - \frac{1}{4} \left( C_{\pi^+\pi^+} +  C_{\pi^-\pi^-} \right) = \frac{1}{2} \left( C_{\pi^0\pi^0}  - C_{\pi^+\pi^+}\right) \, .
\ee
By evaluating the corresponding Wick contractions in the rotated basis, one also finds that the diagrammatic expression of this correlation function is in fact the same as in Eq.\,\eqref{pippi0TM}, but for the sign of the Wilson parameters:
\be
\label{pippimp}
2\, C_{\pi'^+\pi'^-} = C_{\pi^0\pi^0}-C_{\pi^+\pi^+} =  2\, \Delta m^2 \left[\,\, \Maconnpm \,\,\,\, - \,\,\,\, \Madiscpm \,\, \right] \, ,
\ee
The above results could have been anticipated, by observing that the $(u,d)$ and $(u',d')$ fields in the two schemes have exactly the same propagators, in the theory with $\Delta m=0$, in which the correlation functions of Eqs.\,\eqref{pippi0TM} and \eqref{pippimp} are evaluated. The regularization are different, however. The fermion lines in the Feynman diagrams of Eq.\,\eqref{pippimp} represent an $u'$ and a $d'$ quark propagators, which are computed with opposite values of the Wilson parameter ($r_{u'}=+1$ and $r_{d'}=-1$). This is at variance with the diagrams entering Eq.\,\eqref{pippi0TM}, where the fermion lines for the neutral pion propagator represent the same quark field (either $u$ or $d$) and are thus evaluated with the same value of the Wilson parameter.

\begin{figure}[t]
\begin{center}
\includegraphics[width=0.7\textwidth]{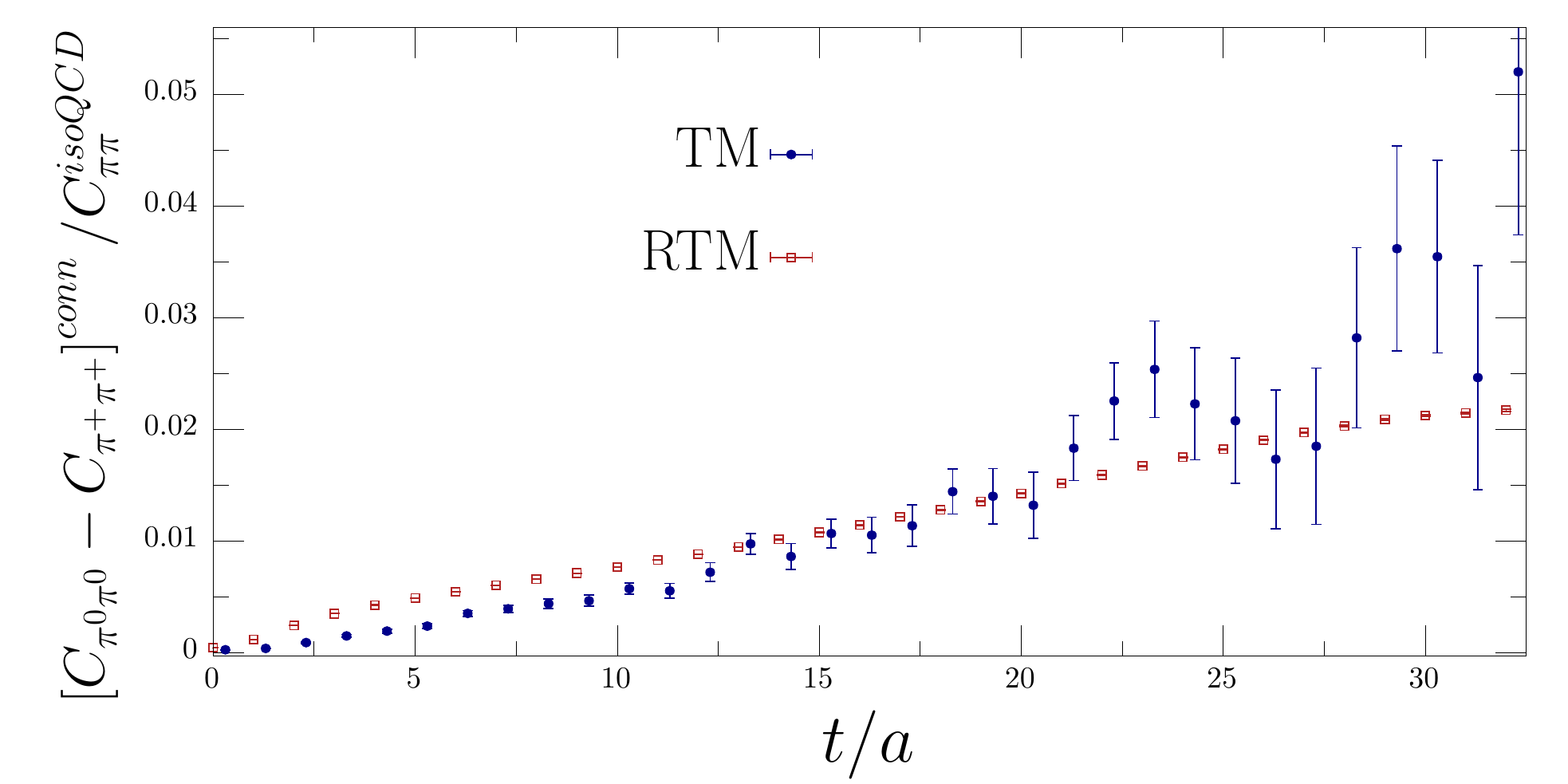}
\includegraphics[width=0.7\textwidth]{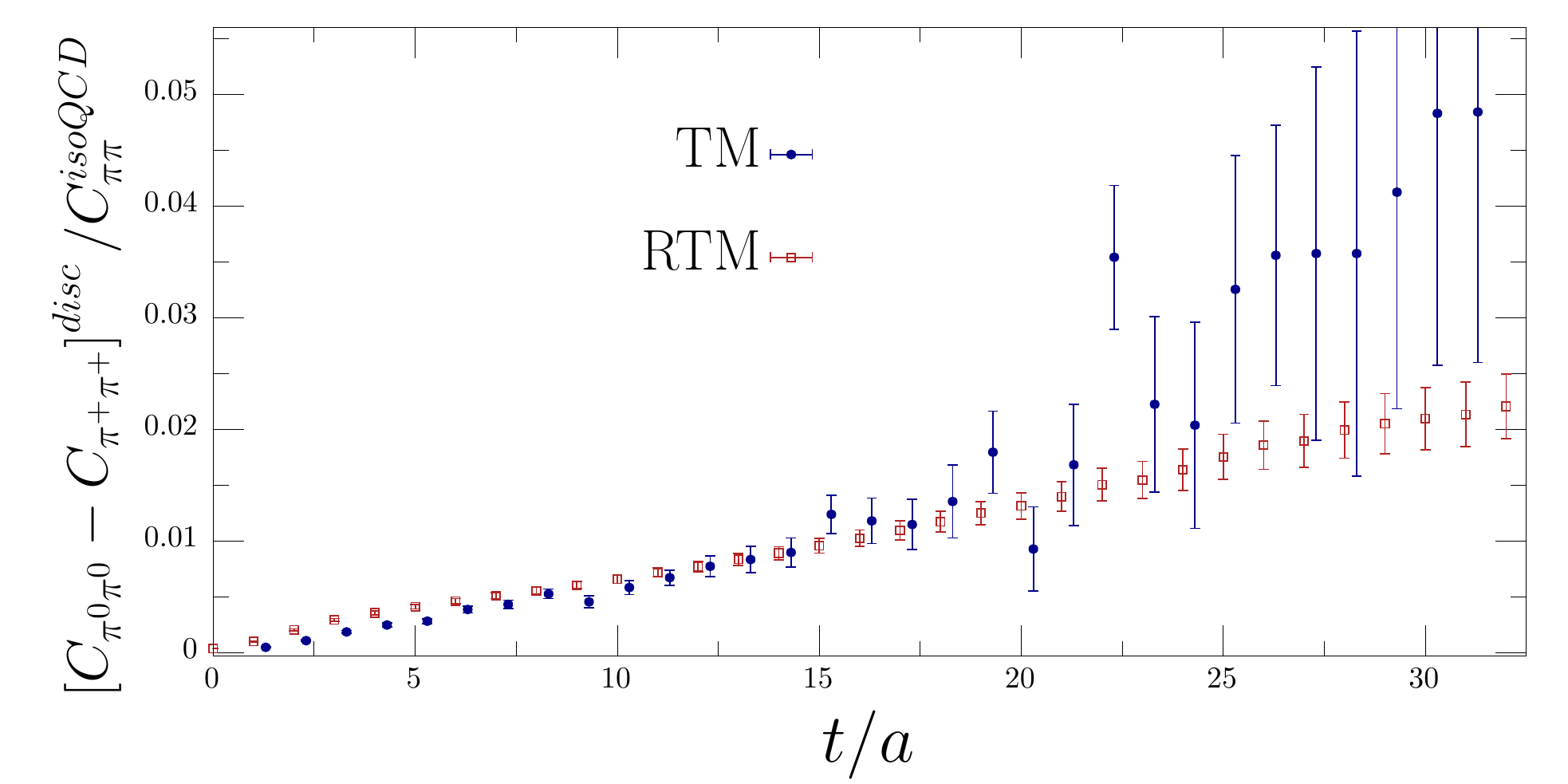}
\vspace{-0.3cm}
\caption{\it 
Results from the connected (top) and disconnected (bottom) quark diagrams contributing to the correlation function $C_{\pi^0\pi^0}-C_{\pi^+\pi^+}$, normalized to the iso-symmetric pion propagator $C_{\pi\pi}^{\rm isoQCD}$, as obtained by using either the standard TM or the RTM scheme. The simulation has been performed on a $L^{3}\times T$ lattice with $L/a=32$ and $T/a=64$, at a value of the lattice spacing $a\simeq 0.094~\rm{fm}$ and of the simulated charged pion mass $M_{\pi}\simeq 260~{\rm MeV}$. A total sample of $N_{\textit{cfg}}=1232$ gauge configurations has been analyzed with a single stochastic source per time slice used to invert the Dirac operator.}
\label{fig:comparison} 
\vspace{-0.5cm}
\end{center}
\end{figure}
In Fig.\,\ref{fig:comparison} we compare the results obtained from the connected (upper plot) and disconnected (lower plot) quark diagrams which contribute to the correlation function $C_{\pi^0\pi^0}-C_{\pi^+\pi^+}$ in the standard TM (Eq.\,\eqref{pippi0TM}) and in the RTM (Eq.\,\eqref{pippimp}) scheme. In order to cancel the leading exponential time behaviour of the correlators, each diagram has been normalized to the corresponding iso-symmetric pion propagator, $C_{\pi\pi}^{\rm isoQCD}$, represented by the single connected diagram without any mass insertions and computed with equal (opposite) values of the Wilson parameter $r$ in the standard TM (RTM) case. The results have been obtained by using one ensemble (cA211.30.32) of the $N_f=2+1+1$ gauge configurations produced with Wilson-clover TM fermions by the ETM Collaboration~\cite{ETMCconfnew}. As it can be seen from the plot, the statistical precision of the correlators evaluated with RTM is significantly improved, in particular for the disconnected quark diagram. 

\section{RTM scheme for isospin breaking QED corrections}

We now discuss the implementation of the RTM regularization in the QCD + QED theory on the lattice, by having in mind the evaluation of IB corrections with the RM123 method for QED \cite{IBem}. While the discussion will proceed along the same lines of the previous section, an additional issue is raised in the QED case by the requirement of ensuring the $U(1)$ gauge invariance on the lattice of the relevant correlation functions.

In the RM123 approach, the QED interactions of quarks is treated as a perturbation, and QED is therefore regularized on the lattice in its non-compact form, see~\cite{IBem} for details. As in the previous section, we postpone the specification of the lattice regularization and start by writing the action in the continuum form.

The photon coupling to the quark fields, defined by the QED covariant derivative, is described by a Lagrangian density that can be written again as the sum of two terms, which are isospin symmetric and isospin violating respectively:
\begin{align}
\label{LQED}
	{\cal L}_{qq\gamma} & = i e\, (q_u \, \bar u \gamma_\mu u + q_d \, \bar d \gamma_\mu d)\, A_\mu = \nn \\
	& = i e\, \frac{q_u + q_d}{2}\, (\bar u \gamma_\mu u + \bar d \gamma_\mu d )\, A_\mu + i e\, \frac{q_u - q_d}{2}\, (\bar u \gamma_\mu u - \bar d \gamma_\mu d )\, A_\mu = \nn \\
	& = i e\, q\, (\bar u \gamma_\mu u + \bar d \gamma_\mu d )\, A_\mu + i e\, \Delta q  \, (\bar u \gamma_\mu u - \bar d \gamma_\mu d )\, A_\mu \nn \\[0.4em]
	& = i e\, q\, (\bar Q \gamma_\mu Q) \, A_\mu + i e\, \Delta q  \, (\bar Q \gamma_\mu \tau_3 Q) \, A_\mu \, .
\end{align}
Here $q_u$ and $q_d$ are the electric charges of the up and down quarks in units of the elementary charge $e$ ($q_u=2/3$ and $q_d=-1/3)$, and 
\be
q = \frac{1}{2}\, (q_u + q_d) \qquad , \qquad \Delta q = \frac{1}{2}\, (q_u - q_d) \ .
\ee
Note that, at variance with the mass case, both terms in the Lagrangian of Eq.\,\eqref{LQED} are treated as perturbations in the RM123 approach of Ref.\,\cite{IBem}. For the purpose of the present discussion, however, in order to define the RTM scheme for QCD+QED, we assume that the term proportional to $\Delta q$ in the action is treated as a perturbation, while for the isospin symmetric term proportional to the average electric charge $q$ no specific assumption is required.

The whole QCD+QED fermionic action can be then written in the form
\be
\label{QEDaction}
{\cal L}_{QCD+QED} = \bar Q \left( \gamma_\mu D^q_\mu + m \right) Q - \Delta m  \, \bar Q \tau_3 Q 
+ i e\, \Delta q  \, (\bar Q \gamma_\mu \tau_3 Q) \, A_\mu \, ,
\ee
where the covariant derivative $D^q_\mu$ contains only the iso-symmetric part of the QED interaction proportional to $q$, while the term proportional to $\Delta q$ has been written out explicitly in Eq.\,\eqref{QEDaction}.

The TM version of the action in Eq.\,\eqref{QEDaction} has the form
\be
\label{TMQEDaction}
{\cal L}_{TM} =  \ \bar Q \left[ \gamma_\mu \widetilde \nabla^q_\mu - i \gamma_5 \tau_3 \, W(m_{cr}) + m \right] Q \, 
-  \, \Delta m\, \bar Q\, \tau_3Q + i e\, \Delta q  \, \left[ (\bar Q \gamma_\mu \tau_3 Q) \, A_\mu \, +  \ldots \right]_{TM} \, ,
\ee
where again the lattice covariant derivative $\nabla^q_\mu$, which enters in the kinetic term but also implicitly in the Wilson term, is defined with $\Delta q=0$. In the last term of order $\Delta q$ the ellipses allude to the occurrence of further, lattice regularization specific interactions terms (see Ref.\cite{IBem} for details). This action is not suitable for direct numerical simulations because, as already noted, for $\Delta m \neq 0$ and $\Delta q \neq 0$ it has a complex fermionic determinant. This issue, however, is avoided in the RM123 approach where the lattice action is expanded in powers of $\Delta m$ and $\Delta q$. For a more complete discussion of a lattice TM regularization of QCD+QED see Ref.\,\cite{TMQED}.

Our motivation, for using the RTM regularization also in the QCD+QED case, is dictated by the remarkable improvement of statistical accuracy. Let us consider, as a numerical example, the QED contribution to the difference $C_{\pi^0\pi^0}-C_{\pi^+\pi^+}$ between the neutral and charged pion propagators, i.e. the same correlation function that we have discussed in the previous section in the evaluation of the strong IB effects. The leading QED contribution to the correlator appears at second order~\cite{IBem}, and it is proportional, as expected, to the charge difference $\Delta q^2$:
\be
\label{pippi0em}
C_{\pi^0\pi^0}-C_{\pi^+\pi^+} = 2\, e^2\, \Delta q^2 \left[ \,\,\Mcconn \,\,\,\,-\,\,\,\, \Mcdisc\,\, \right] \, .
\ee
As in the QCD case, with the standard TM regularization the quark lines in the neutral pion propagator, which gives rise to the disconnected quark diagram contribution in Eq.\,\eqref{pippi0em}, are regularized with the same value of the Wilson parameter $r$. 

At this stage it is important to realize that the difference of correlation functions in Eq.\,\eqref{pippi0em}, which is
in principle defined in QCD+QED and then considered to first order in $e^2$, is proportional, through a coefficient $2\, e^2\Delta q^2$, to a correlation function of the pure iso-symmetric QCD theory (isoQCD). The latter correlation function can hence be evaluated in any convenient lattice regularization of isoQCD, such as the RTM scheme that is advocated here.

Specifically, within any sensible UV-regularization of QCD+QED, the correlation function in Eq.\,\eqref{pippi0em} (at $\Delta m =0$) takes the form
\beqn
\label{pippi0em-th1}
C_{\pi^0\pi^0}-C_{\pi^+\pi^+} &=& 2\, e^2 \Delta q^2 \!\int \! d^4\! y\, d^4\! z\, d{\bf x}\, G_{\mu\nu}(y-z)
\left\langle [P_{\pi^0}(x) P_{\pi^0}^\dagger(0) - P_{\pi^+}(x) P_{\pi^+}^\dagger(0)] 
 \, J_\mu^3(y) J_\nu^3(z) \right\rangle_{\rm isoQCD} \nonumber \\
& & + \, e^2 \! \int \! d^4\!y\, d{\bf x}  \left\langle L_{\rm 1}^{\rm ct}(y) \, \big[P_{\pi^0}(x) P_{\pi^0}^\dagger(0) - P_{\pi^+}(x) P_{\pi^+}^\dagger(0) \big]
 \right\rangle_{\rm isoQCD} \, ,
\eeqn
where $G_{\mu\nu}(y-z)$ is the photon propagator and $e^2 L_{\rm 1}^{\rm ct}$ stands for the standard QCD Lagrangian counterterms that are required in general, at first order in $e^2$, to compensate for the UV-divergencies due to QED interactions. These UV-divergencies arise from contact terms between the two e.m.\ currents. Now, in the particular difference of correlation functions considered here, namely $C_{\pi^0\pi^0}-C_{\pi^+\pi^+}$, the insertion of the $e^2 L_{\rm 1}^{\rm ct}$ counterterms, i.e. the second term in the r.h.s.\ of Eq.~(\ref{pippi0em-th1}), turns out to vanish. This is fairly obvious based on the symmetries of isoQCD, but an explicit diagrammatic proof of the cancellation of the counterterms in the difference between $C_{\pi^0\pi^0}$ and $C_{\pi^+\pi^+}$ is given in sect.\, 5.B of Ref.~\cite{IBem}.

It follows that the correlation function difference $C_{\pi^0\pi^0}-C_{\pi^+\pi^+}$ to first order in $e^2$ is given by simply the first term in the r.h.s.\ of Eq.~(\ref{pippi0em-th1}),  and it is therefore a well defined UV-finite quantity in the pure iso-symmetric QCD theory, once the inserted isotriplet currents and the external pion fields are properly renormalized. Being a well defined and renormalized quantity, we can now also choose to compute the correlator \eqref{pippi0em-th1} using a different UV regularization of the iso-symmetric QCD theory, and a particularly convenient choice is provided again by the RTM scheme.  

In the RTM lattice regularization, the third isospin component of the isotriplet current reads 
\be
\label{J3curr}
J_\mu^3 = \bar Q \gamma_\mu \tau^3 Q =  \bar Q' \gamma_\mu \tau^1 Q' = J_\mu^{\prime 1} \, ,
\ee 
once it is expressed in terms of the quark fields $Q'=(u',d')$ of the rotated basis \eqref{qrotated}, which are now regularized with $r_{u'}=-r_{d'}=+1$. Moreover, Eq.\,\eqref{cpippim} tells us that we have the
equality $P_{\pi^0}(x) P_{\pi^0}^\dagger(0) - P_{\pi^+}(x) P_{\pi^+}^\dagger(0) = 2 P_{\pi^{\prime +}}(x) P_{\pi^{\prime -}}^\dagger(0)$, where, in the notation of Eq.\,\eqref{pirotated}, $P_{\pi^{\prime +}} = P_{\pi^{\prime -}}^\dagger = \bar d' \gamma_5 u'$ . Hence we arrive at the result 
\be
\label{respippimp}
C_{\pi^0\pi^0}-C_{\pi^+\pi^+} = 2\, C_{\pi'^+\pi'^-} =  2 e^2 \Delta q^2 Z_A^2 \!\int \! d^4\! y\, d^4\! z\, d{\bf x}\, G_{\mu\nu}(y-z) \left\langle P_{\pi^{\prime +}}(x) P_{\pi^{\prime -}}^\dagger(0)
 \, J_\mu^{\prime 1}(y) J_\nu^{\prime 1}(z) \right\rangle_{\rm isoQCD} \, ,
\ee
which tells us that the difference $C_{\pi^0\pi^0}-C_{\pi^+\pi^+}$ is proportional to the correlation function $C_{\pi'^+\pi'^-}$ in the RTM regularization of the iso-symmetric QCD theory. In Eq.\,\eqref{respippimp} we have also included explicitly the normalization factor $Z_A^2$, where $Z_A$ is the (finite) QCD renormalization constant of the local Wilson axial-vector current, which properly normalizes, in the RTM scheme, the local vector current $J_\mu^{\prime 1}$. By performing in Eq.\,\eqref{respippimp} the relevant Wick contractions we then find
\be
\label{pippimpem}
C_{\pi^0\pi^0}-C_{\pi^+\pi^+} = 2\, C_{\pi'^+\pi'^-} = 2\, e^2\, \Delta q^2 \left[\,\, \Mcconnpm \,\,\,\, - \,\,\,\,  \Mcdiscpm \,\, \right] \, ,
\ee
where now, however, at variance with Eq.\,\eqref{pippi0em}, the fermion lines in both diagrams represent a $u'$ and a $d'$ quarks, and are regularized, therefore, with opposite values of the Wilson parameter.

\begin{figure}[t]
\begin{center}
\includegraphics[width=0.9\textwidth]{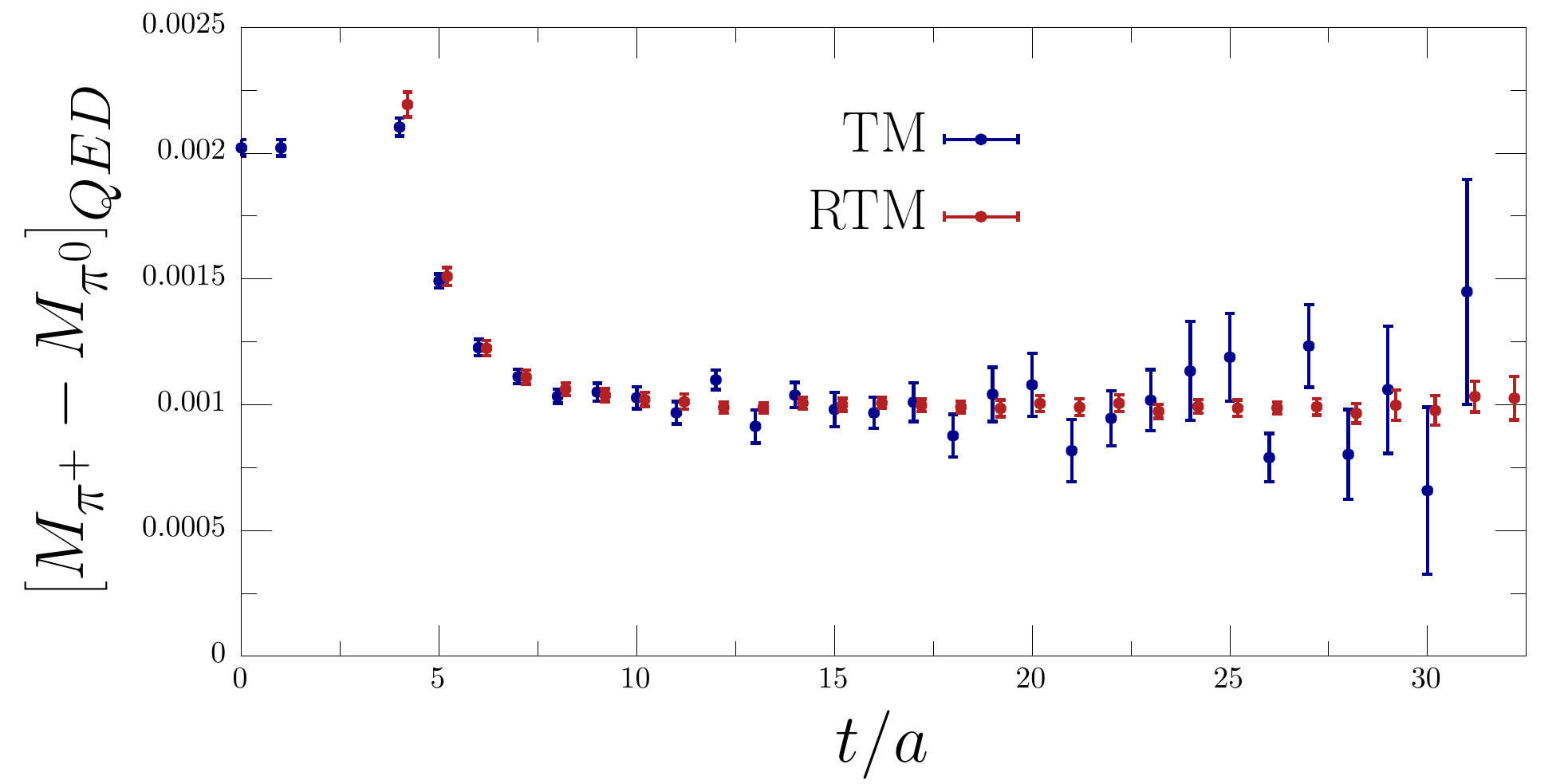}
\vspace{-0.3cm}
\caption{\it \label{fig.comparisonsQED}
Effective mass plot for the QED contribution to the mass difference $M_{\pi^+} -M_{\pi^0} $ at ${\cal O}(\Delta q^2)$ as obtained by using either the standard TM or the RTM scheme. The simulation has been performed on a $L^{3}\times T$ lattice with $L/a=32$ and $T/a=64$, at a value of the lattice spacing $a\simeq 0.089~\rm{fm}$ and of the simulated charged pion mass $M_{\pi}\simeq 320~{\rm MeV}$. A total sample of $N_{\textit{cfg}}=100$ gauge configurations has been analyzed with a single stochastic source per time slice used to invert the Dirac operator.}
\vspace{-0.5cm}
\end{center}
\end{figure}

A comparison of the results obtained for the pion mass splitting at ${\cal O}(\Delta q^2)$ by using either the standard TM or the RTM scheme is shown in Fig.\,\ref{fig.comparisonsQED}. In both cases, the pion mass difference $[M_{\pi^{+}}- M_{\pi^{0}}]_{QED}$ has been extracted using
\be
\left[M_{\pi^{+}}- M_{\pi^{0}}\right]_{QED} = 2e^{2}\Delta q^{2}\partial_{t}\frac{\Mcconnnor \,\,\,\, -\,\,\,\,  \Mcdiscnor      }{\Isocorr}~,
\ee
where the diagram in the denominator is the iso-symmetric pion propagator $C_{\pi\pi}^{\rm isoQCD}$, and on each quark line appearing in the previous diagrams the sign of the Wilson parameter $r$ has been chosen appropriately according to whether standard TM or the RTM scheme has been considered. The operator $-\partial_{t}$ in the previous equation is defined through
\begin{align}
\label{eq:def_meff}
-\partial_{t} \frac{\delta C(t)}{C(t)} &= \frac{1}{F(T/2 -t ,M)}\left(\frac{\delta C(t)}{C(t)} -\frac{\delta C(t-1)}{C(t-1)}\right) ~ ,
\end{align}
where $M$ is the ground state mass extracted from the correlator $C(t)$, $\delta C(t)$ is the correlator difference in Eq.~(\ref{pippimp}) or in Eq.~(\ref {pippimpem}), $T$ is the temporal extent of the lattice and $F(x,M)$ is given by
\begin{align}
F(x,M) &= x\tanh{(Mx)} - (x+1)\tanh{(M(x+1))} ~ .
\end{align}
The results have been obtained by using one ensemble (A40.32) of the $N_f=2+1+1$ gauge configurations produced with Wilson TM fermions by the ETM Collaboration~\cite{ETMCconfold}.
We see from the plot that, similarly to the QCD case, the statistical accuracy of the lattice results obtained with the RTM scheme is significantly improved with respect to the standard TM case.

We conclude this section with a couple of remarks concerning the finiteness and gauge invariance of the correlation function $C_{\pi'^+\pi'^-}$ of Eq.\,\eqref{respippimp}. Since this correlator is just a convenient regularization in the pure QCD iso-symmetric theory of a difference of renormalized and gauge invariant correlators defined in the full QCD+QED theory, which is represented for instance by Eq.\,\eqref{TMQEDaction} or any other regularization of QCD+QED, we know that it must be both finite and gauge invariant, provided the electromagnetic currents and the external fields entering in Eq.\,\eqref{respippimp} have been properly renormalized in the RTM scheme. In the specific case of the correlator\,\eqref{respippimp}, it can be easily checked, by inspection, that additional contact terms generated by the double insertion of the electromagnetic current are indeed absent. This follows from the fact that the operator $(\bar u' \gamma_\mu d')^2$, which represents the relevant double insertion of the current $J_\mu^{\prime 1} = \bar u' \gamma_\mu d' + \bar d' \gamma_\mu u'$ in the correlation function\,\eqref{respippimp}, has a flavor-changing stucture ($\Delta u' = - \Delta d'$=2) such that it cannot mix with lower-dimensional operators.

Similarly, one can also verify that the result of Eq.\,\eqref{respippimp}, or equivalently Eq.\,\eqref{pippimpem}, is actually independent of the gauge fixing choice made for the photon propagator. This follows from the fact that in the iso-symmetric QCD theory, where the correlation function is going to be evaluated, the current coupled to the photon propagator in Eq.\,\eqref{respippimp} is conserved. Formally, in this theory one has
\be
\partial_\mu (\bar Q' \gamma_\mu \tau_1 Q') =  \left[\bar Q' \left( \overleftarrow{D}_\mu \gamma_\mu + \overrightarrow{D}_\mu \gamma_\mu \right) \tau_1 Q' \right] = (m - m)  \left( \bar Q'  \tau_1 Q' \right) = 0\, .
\ee
On the lattice, where current conservation holds up to immaterial (in our case O($a^2$)) cutoff artifacts, one has thus to employ the proper conserved and chiral covariant current $J_\mu^{\prime 1}$, which, as already mentioned, in the RTM scheme, is the bare local current times the UV finite normalization factor $Z_A$.

\section{Conclusions}

In this paper we have pointed out that a scheme of lattice twisted-mass fermion regularization that was previously introduced to make feasible unquenched simulation with mass non-degenerate quarks can be very conveniently employed for calculations of the leading IB corrections to mesonic observables both in pure QCD and in QCD+QED. The scheme, that we denote as {\it rotated twisted-mass} (RTM) scheme, is in fact suitable for being implemented together with the RM123 approach, in which the IB terms of the action are treated as a perturbation.
The main advantage of the RTM scheme is that, by enabling the evaluation of the mesonic observables of interest through lattice correlation functions in which quark and antiquark fields are regularized with opposite values of the Wilson parameter, it leads to a significant improvement of the statistical accuracy. 

In order to show the efficiency of the proposed approach, we have applied the RTM scheme to compute the charged-neutral pion mass splitting both in QCD and in QCD+QED at ${\cal O}(\Delta m^2)$ and ${\cal O}(\Delta q^2)$ respectively, at a fixed value of simulated quark masses and lattice spacing. The results have been compared with the corresponding results obtained by using the standard TM regularization and the improvement achieved with the RTM scheme is manifest.

We plan to apply this proposal to more complete lattice calculations, including the evaluation of the pion mass splitting at physical values of the quark masses and in the continuum and infinite volume limits. Moreover, we want to explore the effectiveness of the same approach 
in the calculations of 3-point functions, like for instance those relevant for kaon or pion semileptonic decays, as well as to the calculation of IB baryonic observables, like the neutron-proton mass splitting.

\section*{Acknowledgements}

We thank G. Martinelli  for stimulating discussion. We acknowledge CINECA for the provision of CPU time under the specific initiative INFN-LQCD123. F.S. G.G and S.S. are supported by the Italian Ministry of University and Research (MIUR) under grant PRIN20172LNEEZ. F.S. and G.G are supported by INFN under GRANT73/CALAT.



\end{document}